\documentclass[preprint,aps,showpacs]{revtex4}
\usepackage{graphicx,epsf}
\usepackage{lscape}
\usepackage{citesort}
\usepackage[psamsfonts]{amssymb}

\begin{document}

\title{{\Large{Semileptonic decays of $B (B_s)$ to light tensor mesons }}}

\author{\small R. Khosravi\footnote {e-mail: rezakhosravi @ cc.iut.ac.ir}, S.
Sadeghi }

\affiliation {Department of Physics, Isfahan University of
Technology, Isfahan 84156-83111, Iran }

\begin{abstract}
The semileptonic  $B_s (B) \to K_2^*(a_2, f_2)
\ell\nu$, $\ell=\tau, \mu$  transitions are investigated in the frame work of the
three-point QCD sum rules. Considering the
quark condensate contributions, the relevant form factors  of these
transitions are estimated. The branching ratios of these channel
modes are also calculated at different values of the continuum
thresholds of the tensor mesons and compared with the obtained data
for other approaches.
\end{abstract}


\maketitle

\section{Introduction}
Investigation of the $B$ meson decays into tensor mesons are useful
in several aspects such as CP asymmetries, isospin symmetries and
the longitudinal and transverse polarization fractions. A
large isospin violation has already been experimentally detected in
$B\to\omega K_2^*(1430)$ mode \cite{Aubert1}. Also, the decay mode
$B\to\phi K_2^*(1430)$ is mainly dominated by the longitudinal
polarization \cite{Aubert2,Aubert3}, in contrast with the $B\to \phi
K^*$ where the transverse polarization is comparable with the
longitudinal one \cite{Barberio}. Therefore, nonleptonic and
semileptonic decays of $B$ meson can play an
important role in the study of the particle physics.

In the flavor $SU(3)$ symmetry, the light $p$-wave tensor
mesons with $J^P=2^+$ containing iso-vector mesons  $a_2(1320)$,
iso-doublet states  $K^*_2 (1430)$,  and two iso-singlet mesons
$f_2(1270)$ and $f'_2 (1525)$, are building the ground state nonet
which have been experimentally established \cite{Wang,Dombrowski}.
The quark content $q\bar{q}$ for the iso-vector and iso-doublet tensor
resonances are obvious. The iso-scalar tensor states, $f_2(1270)$ and
$f'_2(1525)$ have a mixing wave functions where mixing angle should
be small \cite{PDG1,Li}. Therefore, $f_2(1270)$ is primarily a
$(u\bar{u}+d\bar{d})/\sqrt{2}$ state, while $f'_2(1525)$ is
dominantly $s\bar{s}$ \cite{Cheng}.

As a nonperturbative method, the QCD sum rules is a well established
technique in the hadron physics since it is based on the fundamental
QCD Lagrangian. The semileptonic decays of $B$ to the light mesons
involving $\pi$, $K (K^*, K_{0}^{*})$, and $a_1$ have been studied
via the three-point QCD sum rules (3PSR), for instance $B\to \pi
\ell \nu$ \cite{Ball1}, $B\to K \ell^+ \ell^-$, $B\to K^{*} \ell^+
\ell^-$ \cite{ColFa,CoDom}, $B_s\to K^*_0 \ell \nu$ \cite{Yang2},
$B_s\to (K_0^*, f_0) \ell^+ \ell^-$ \cite{Khosravi1} and $B\to a_1
\ell^+ \ell^-$ \cite{Khosravi2}. The determination of the form
factor value $T_1(0) = 0.35\pm 0.05$ relevant for the $B \to K^*
\gamma$ and $B \to K^* \ell^+ \ell^-$ \cite{CoDom,Ball3} decays
allowed to predict the ratio $\Gamma(B\to K^* \gamma)/\Gamma(b\to s
\gamma) = 0.17\pm 0.05$, which agrees with the experimental
measurements \cite{CLEO1,CLEO2,ALEPH}. The obtained results of the
decay $B\to \pi \ell \nu$ \cite{Ball1}, and simulations on the
lattice \cite{Abada,Allton,UKQCD} are in a reasonable agreement.

However, considering the structure of the 3PSR for heavy-light
transitions shows a difficulty which is due to the dependence of the
various terms of the short-distance expansion to mass of the
$b$-quark \cite{PCol}. Therefore, some authors claimed that the 3PSR
is not a well-established tool for heavy-light transition, and is
ill-behaved for large $m_b$ mass which has been discussed in
\cite{Ali}.

It should be noted that the treatment of the 3PSR and light-cone QCD
sum rules (LCSR) are different for the light hadron in the final
state. The 3PSR, without considering wave functions, introduces a
three-point correlation function with appropriate interpolating
currents and extract the perturbative and nonperturbative
contributions of the transition form factors. In the limit of
$m_b\to \infty$, the coefficients of the nonperturbative effects
such as quark-quark and quark-gluon condensate are increased with
$m_b$ faster than the perturbative coefficient. In the light-cone
sum rules (LCSR) method,  this problem does not appear since the
nonperturbative effects are included in the hadron distribution
amplitudes \cite{Ball4}.

This problem could be irrelevant for the actual value of the
$b$-quark mass, and also for particular processes and final states.
For instance, the obtained results based on the 3PSR for the $B\to
\pi$, and $B\to K^{*}$ transitions are in good agreement with the
experimental values, as it was already mentioned. On the other hand,
using the Borel transformation exponentially suppresses the
contributions of the highest-order operators. However, it is
necessary to compare the results obtained for heavy-light transition
from the 3PSR with other methods, especially the LCSR and
experimental data if exist.

It should be emphasized that a suitable choice of the Borel parameter
interval keeps the convergence of the condensate expansion under
control. Therefore, neglecting the higher-dimensional terms do not
introduce a large error. In Ref. \cite{Ali}, authors have
obtained  $T^{B\to K^*}_1(0)\simeq0.5 - 0.6$ and $0.32$ via the 3PSR
and the LCSR, respectively, where the 3PSR value is larger than that
for the LCSR.  Also, this quantity has been calculated $0.38$ based
on the 3PSR in \cite{Ball5}, which has a reasonable agreement
with the LCSR value. The main reason for the difference between
results based on the 3PSR in the two papers is due to the difference in
selection of the Borel parameter interval \cite{Ali}.

In this work, we investigate  the $B (B_s)\to K_2^*(a_2, f_2) \ell
\nu$ decays within the 3PSR method. For analysis of these decays,
the form factors and their branching ratio values are calculated. So
far, the form factors of the semileptonic decays $B (B_s)\to
K_2^*(a_2, f_2) \ell \nu$ have been studied via different approaches
such as the LCSR \cite{Yang}, the perturbative QCD (PQCD)
\cite{Wang}, the large energy effective theory (LEET)
\cite{Charles,Ebert,Datta}, and the ISGW II model \cite{Scora}. A
comparison of our results for the form factor values in $q^2=0$ and
branching ratio data with predictions obtained from other
approaches, especially the LCSR, is also made.

The plan of the present paper is as follows: The 3PSR approach for
calculation of the relevant form factors of the $B (B_s)\to
K_2^*(a_2, f_2) \ell\nu$ decays presented in Section II. In the
final section, the value of the form factors in $q^2=0$ and the
branching ratio of the considered decays are reported. For a better
analysis, the form factors and differential branching ratios related
to these semileptonic decays are plotted with respect to the
momentum transfer squared $q^2$.

\section{Theoretical framework}
In order to study of $B (B_s)\to K_2^*(a_2, f_2) \ell \nu$ decays, we focus on the exclusive decay $B_s\to K^*_2$  via the 3PSR.
The $B_s\to K^*_2 \ell\nu$ decay governed by the tree level $b\to u
$ transition (see Fig. \ref{F1}).
\begin{figure}[th]
\vspace{0cm}\label{F1}
\includegraphics[width=4cm,height=4cm]{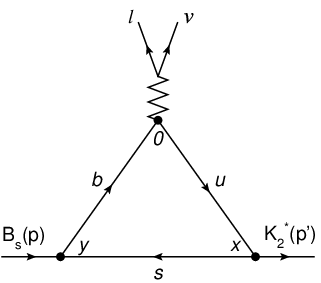}
\hspace{0.5cm}
\caption{Schematic picture of the spectator mechanism for the $B_{s}\rightarrow K_{2}^{*} \ell \nu$ decay.}\label{F1}
\end{figure}
In the framework of the 3PSR, the first step is appropriate definition of correlation function. In this work, the correlation function should be taken as
\begin{eqnarray}\label{eq21}
\Pi_{\alpha\beta\mu}(p^2, p'^2, q^2)= i\int \int d^{4}x~d^{4}y~
e^{i(p'x-py)} \left\langle0\left| {\cal T}\left\{
j^{K_{2}^{*}}_{\alpha\beta}(x) j_{\mu}(0) {j^{B_{s}}}(y)\right\}
\right|0\right\rangle,
\end{eqnarray}
where $p$ and $p'$ are four-momentum of the initial and final mesons, respectively. $q^2$ is the squared momentum transfer and $\cal T$ is the time ordering operator. $j_{\mu}=\bar{u}\gamma_{\mu} (1-\gamma_{5}) b$ is the transition current. $j^{B_s}$ and $j^{K_{2}^{*}}_{\alpha\beta}$ are also the interpolating currents of the $B_s$ and the tensor meson $K_{2}^{*}$, respectively. With considering all quantum numbers, their interpolating currents can be written as
\begin{eqnarray}\label{eq22}
j^{B_s}(y)&=&\bar b(y) \gamma_{5} s(y),\nonumber\\
{j^{K_{2}^{*}}_{\alpha \beta }}(x) &=& \frac{i}{2}\left[ \bar s(x) \gamma _\alpha
\stackrel\leftrightarrow D_\beta(x) u(x) + \bar s(x) \gamma _\beta
\stackrel\leftrightarrow D_\alpha(x) u(x)\right],
\end{eqnarray}
where $\stackrel\leftrightarrow D_\mu(x)$ is the four-derivative
with respect to $x$ acting at the same time on the left and right.
It is given as
\begin{eqnarray*}\label{eq23}
\stackrel\leftrightarrow D_\mu(x)& =&
\frac{1}{2}\left[\stackrel\rightarrow D_\mu(x) - \stackrel\leftarrow
D_\mu(x)\right],\nonumber\\
\stackrel\rightarrow D_\mu(x) &=& \stackrel\rightarrow\partial_\mu(x) - i \frac{g}{2} \lambda^a \textbf{\emph{A}}^a_\mu(x), \\
\stackrel\leftarrow D_\mu(x)&=&\stackrel\leftarrow\partial_\mu(x)+ i
\frac{g}{2} \lambda^a \textbf{\emph{A}}^a_\mu(x),
\end{eqnarray*}
where $\lambda^{a}$ and $\textbf{\emph{A}}^a_\mu(x)$ are the
Gell-Mann matrices and the external gluon fields, respectively.

The correlation function is a complex function of which the imaginary
part comprises the computations of the phenomenology and real part
comprises the computations of the theoretical part (QCD). By linking these two parts
via the dispersion relation, the physical quantities  are calculated.
In the phenomenological part of the QCD sum rules approach, the correlation function in Eq. (\ref{eq21}) is calculated by inserting
two complete sets of intermediate states with the same quantum numbers as $B_s$ and $K_{2}^{*}$. After performing four-integrals over $x$ and $y$, it will be:
\begin{eqnarray}\label{eq24}
\Pi_{\alpha\beta\mu}=-\frac{\langle0|j^{K_{2}^{*}}_{\alpha\beta}|K_{2}^{*}(p')\rangle
\langle K_{2}^{*}(p')|j_{\mu}|B_{s}(p)\rangle \langle
B_{s}(p)|j^{B_{s}}|0\rangle}{(p^{2}-m_{B_{s}}^{2})(p'^{2}-m_{K_{2}^{*}}^{2})}+\mbox{higher
states}.
\end{eqnarray}
In Eq. (\ref{eq24}), the vacuum to initial and final meson state matrix elements are defined as
\begin{eqnarray}\label{eq25}
\langle0|j^{K_{2}^{*}}_{\alpha\beta}|K_{2}^{*}(p',\varepsilon)\rangle = f_{K_{2}^{*}} m_{K_{2}^{*}}^{2}\varepsilon_{\alpha\beta},~~~~~~~~~
\langle 0|j^{B_{s}}|B_{s}(p)\rangle =- i\frac{f_{B_{s}}m_{B_{s}}^{2}}{(m_{b}+m_{s})},
\end{eqnarray}
where $f_{K_{2}^{*}}$ and $f_{B_s}$ are the leptonic decay constants of $K_{2}^{*}$ and $B_s$ mesons, respectively.
$\varepsilon_{\alpha\beta}$ is polarization tensor of $K_{2}^{*}$.
The transition current give a contribution to these matrix elements and it can
be parametrized in terms of some form factors using the
Lorentz invariance and parity conservation. The correspondence between a vector meson and a tensor meson
allows us to get these parametrization in a comparative way ( for more information see \cite{Wang}). The parametrization of $B\to T$ form
factors is analogous to the $B\to V$ case except that the $\varepsilon$ is replaced by $\varepsilon_{T}$, as follows:
\begin{eqnarray}\label{eq26}
\lefteqn{c_V
\langle K_{2}^{*}(p',\varepsilon)|\bar u \gamma_{\mu}(1-\gamma_5)
b|B_s(p)\rangle  =-i \varepsilon_{T\mu}^{*}(m_{B_s}+m_{K_{2}^{*}})A_1(q^2) +i
(p+p')_{\mu}(\varepsilon_T^{*}. q)\frac{A_2(q^2)}{ m_{B_s}+m_{K_{2}^{*}}}}\hspace*{2.8cm}\nonumber\\
&& {}+iq_\mu (\varepsilon_T^{*}.q) \frac{2 m_{K_{2}^{*}}}{q^2} \left(A_3(q^2)-A_0(q^2)\right)
+\epsilon_{\mu\nu\rho\sigma}\varepsilon_T^{*\nu} p^{\rho}p'^{\sigma}\frac{2V(q^2)}{m_{B_s}+m_{K_{2}^{*}}}\hspace*{0.5cm}\\
{\rm with\ }A_3(q^2) &=& \frac{m_{B_s}+m_{K_2^*}}{2m_{K_2^*}}A_1(q^2)-\frac{m_{B_s}-m_{K_2^*}}{2m_{K_2^*}}A_2(q^2)\mbox{~~and~~}
A_0(0) =  A_3(0),
\end{eqnarray}
where $q=p-p'$, $P=p+p'$, and $\varepsilon^{*}_{T\mu}=\frac{p_{_\lambda}}{m_{B_s}}\varepsilon_{\mu\lambda}$.
The factor $c_V$ accounts for the flavor content of particles: $c_V = \sqrt{2}$
for $a_2$, $f_2$ and $c_V = 1$ for $K^{*}_2$ \cite{Ball}. Inserting
Eqs. (\ref{eq25}) and (\ref{eq26}) in Eq. (\ref{eq24}) and
performing summation over the polarization tensor as
\begin{eqnarray*}\label{eq27}
\varepsilon_{\mu\nu}\varepsilon_{\alpha\beta}&=&\frac{1}{2}T_{\mu\alpha}T_{\nu\beta}+\frac{1}{2}T_{\mu\beta}T_{\nu\alpha}-\frac{1}{3}T_{\mu\nu}T_{\alpha\beta},
\end{eqnarray*}
where $T_{\mu\nu}=-g_{\mu\nu}+\frac{p'_{\mu}p'_{\nu}}{m_{K_{2}^{*}}^{2}}$, the final representation of the physical side is obtained as
\begin{eqnarray}\label{eq28}
\Pi_{\alpha\beta\mu}&=&\frac{f_{B_s}m_{B_s}}{(m_{b}+m_{s})}\frac{f_{K_{2}^{*}}m_{K_{2}^{*}}^{2}}{(p^2-m_{B_s}^2)(p'^2-m_{K_{2}^{*}}^2)}\left\{ V'(q^2)
i\epsilon_{\beta\mu\rho\sigma}p_{\alpha}p^{\rho}p'^{\sigma}+A'_0(q^2) p_{\alpha}p_{\beta}p'_{\mu}\right.\nonumber\\
&+&\left.A'_1(q^2) g_{\beta\mu}p_{\alpha}+A'_2(q^2) p_{\alpha}p_{\beta}p_{\mu}\right\}+\mbox{higher states}.
\end{eqnarray}
For simplicity in calculations, the following
redefinitions have been used in Eq. (\ref{eq28}):
\begin{eqnarray*}\label{eq29}
V'(q^2)&=&\frac{V(q^2)}{m_{B_s}+m_{K_{2}^{*}}},~~~~~~~~~~~~~~~~~~~~ A'_0(q^2)=-\frac{ m_{K_{2}^{*}} (A_3(q^2)-A_0(q^2))}{ q^2},\nonumber\\
A'_1(q^2)&=&-\frac{(m_{B_s}+m_{K_{2}^{*}})}{2}A_1(q^2),~~~~~~~~~
A'_2(q^2)=\frac{A_{2}(q^2) }{2(m_{B_s}+m_{K_{2}^{*}})}.
\end{eqnarray*}
Now, the QCD part of the correlation function is calculated by expanding it in terms of the OPE at large negative value of $q^{2}$:
\begin{eqnarray}\label{eq210}
\Pi_{\alpha\beta\mu}= C^{(0)}_{\alpha\beta\mu}{\rm I}+C^{(3)}_{\alpha\beta\mu}\langle0|\bar{\Psi}\Psi|0\rangle+C^{(4)}_{\alpha\beta\mu}\langle0|G^{a}_{\rho\nu}G_{a}^{\rho\nu}|0\rangle+C^{(5)}_{\alpha\beta\mu}\langle 0| \bar{\Psi} \sigma_{\rho\nu} T^a G_a^{\rho\nu} \Psi| 0\rangle +...,
\end{eqnarray}
where $C^{(i)}_{\alpha \beta \mu}$ are the Wilson coefficients, I is the unit operator, $\bar{\Psi}$ is the local fermion field operator and $G^a_{\rho\nu}$ is the gluon strength tensor. In  Eq. (\ref{eq210}) the first term is contribution of the perturbative and the other terms are contribution of the non-perturbative part.

To compute the portion of the perturbative part (Fig. \ref{F1}),
using the Feynman rules for the bare loop, we obtain:
\begin{eqnarray}\label{eq211}
C^{(0)}_{\alpha\beta\mu}&=&-\frac{i}{4}\int \int d^4x~ d^4y~ e^{i(p'x -py)}\left\{{\rm Tr}\left[S_{s}(x-y)\gamma_{\alpha}\stackrel\leftrightarrow D_\beta(x)S_{u}(-x)\gamma_{\mu}(1-\gamma_{5})S_{b}(y)\gamma_{5}\right]\right.\nonumber\\ &+& \left. {\rm Tr}[\alpha\leftrightarrow\beta]\right\},
\end{eqnarray}
taking the partial derivative with respect to $x$ of the
quark free propagators, and performing the Fourier
transformation and using the Cutkosky rules, i.e.,
$\frac{1}{p^2-m^2}\rightarrow -2i\pi\delta(p^2-m^2)$,
imaginary part of the $C^{(0)}_{\alpha\beta\mu}$ is calculated as
\begin{eqnarray}\label{eq212}
{\rm Im} \left[C^{(0)}_{\alpha\beta\mu}\right]&=&\frac{1}{8\pi}\int d^{4}k \delta(k^2-m_s^2)\delta((p + k)^2- m_b^2)\delta((p' + k)^2- m_u^2)(2k+p')_{\beta}\nonumber\\
&\times&{\rm Tr}\left[(\not k +m_{s})\gamma_{\alpha}(\not p'+\not k +m_{u})\gamma_{\mu}(1-\gamma_{5})(\not p+\not k +m_{b})\gamma_{5}\right]
+\{\alpha \leftrightarrow \beta\},
\end{eqnarray}
where $k$ is four-momentum of the spectator quark $s$. To solve
the integral in Eq. (\ref{eq212}), we will have to deal with the
integrals such as $I_0$, $I_{\alpha}$,  $I_{\alpha\beta}$ and $I_{\alpha\beta\mu}$ with respect to $k$. For example $I_{\alpha\beta\mu}$ can be as:
\begin{eqnarray*}
I_{\alpha\beta\mu}(s, s',q^2) =\int d^4k~[k_\alpha k_\beta k_\mu]
\delta(k^2-m_s^2)\delta((p + k)^2- m_b^2)\delta((p' + k)^2- m_u^2).
\end{eqnarray*}
where $s=p^2$ and $s'=p'^2$.
$I_0$, $I_{\alpha}$,  $I_{\alpha\beta}$ and $I_{\alpha\beta\mu}$ can be taken as an appropriate tensor structure as follows:
\begin{eqnarray}\label{eq213}
I_0&=&\frac{1}{4\sqrt{\lambda(s, s', q^2)}} ,\nonumber \\
I_\alpha&=& B_1 [p_\alpha]+ B_2 [p'_\alpha], \nonumber \\
I_{\alpha\beta}&=& D_1 [g_{\alpha\beta}] +D_2  [p_\alpha p_\beta]+D_3 [p_\alpha p'_\beta+p'_\alpha p_\beta]+D_4 [p'_\alpha p'_\beta], \nonumber\\
I_{\alpha\beta\mu} &=& E_{1}[g_{\alpha\beta}p_{\mu}+g_{\alpha\mu}p_{\beta}+g_{\beta\mu}p_{\alpha}]+E_{2}[g_{\alpha\beta}p'_{\mu}+g_{\alpha\mu}p'_{\beta}+g_{\beta\mu}p'_{\alpha}]
+E_3[p_{\alpha}p_{\beta}p_{\mu}]\nonumber\\
&+&E_4[p'_{\alpha}p_{\beta}p_{\mu}+p_{\alpha}p'_{\beta}p_{\mu}+p_{\alpha}p_{\beta}p'_{\mu}]+E_5[p'_{\alpha}p'_{\beta}p_{\mu}+p'_{\alpha}p_{\beta}p'_{\mu}+p_{\alpha}p'_{\beta}p'_{\mu}]\nonumber\\&+&E_6[p'_{\alpha}p'_{\beta}p'_{\mu}],
\end{eqnarray}
The quantities $\lambda(s, s', q^2)$, $B_l~ (l=1, 2)$, $D_j~ (j=1,...,4)$, and $E_r~(r=1,...,6)$, are indicated in the Appendix.
Using the relations in Eq. (\ref{eq213}), ${\rm Im} [C^{(0)}_{\alpha\beta\mu}]$ can
be calculated for the each structure corresponding to Eq. (\ref{eq28}) as follows:
\begin{eqnarray}\label{eq214}
{\rm Im}\left[C^{(0)}_{\alpha\beta\mu}\right]&=&  \rho_{_V}
(i\epsilon_{\beta\mu\rho\sigma}p_{\alpha}p^{\rho}p'^{\sigma})+\rho_{_0}(p_{\alpha}p_{\beta}p'_{\mu})+\rho_{_1}(g_{\beta\mu}p_{\alpha})
+\rho_{_2}(p_{\alpha}p_{\beta}p_{\mu}).
\end{eqnarray}
where the spectral densities $\rho_{i},~ (i=V, 0, 1, 2)$ are found as
\begin{eqnarray*}\label{eq215}
\rho_{_V}(s,s',q^{2})&=&24 B_1 \sqrt{\lambda}\left[B_{1}(m_{s}-m_{b})+B_{2}(m_{s}-m_{u})+m_{s}I_{0}\right],\nonumber\\
\rho_{_0}(s,s',q^2)&=&12[D_{2}(m_{s}-m_{b})+D_{3}(m_{s}-m_{u})+2B_{1}m_{s}-2E_{4}(m_{b}-m_{s})],\nonumber\\
\rho_{_1}(s,s',q^2)&=&3B_{1}[2m_s^2(m_{b}+m_{u}-m_s)-m_{s}(2m_{b}m_{u}+u)+\Delta( m_{s}-m_u)+\Delta'(m_{s}-m_{b})]\nonumber\\
&+&6D_{1}(m_{s}-m_{u})-24E_{1}(m_{b}-m_{s}),\nonumber\\
\rho_{_2}(s,s',q^2)&=&24[D_{2}m_{s}+E_{3}(m_{s}-m_{b})].
\end{eqnarray*}
Using the dispersion relation, the
perturbative part contribution of the correlation function can
be calculated as follows:
\begin{equation}\label{eq216}
C^{(0)}_{i}=\int ds'\int ds \frac{\rho_{i}(s,s',q^{2})}{(s-p^{2})(s'-p'^{2})}.
\end{equation}

\begin{figure}[th]
\vspace{0cm}
\includegraphics[width=6cm,height=3cm]{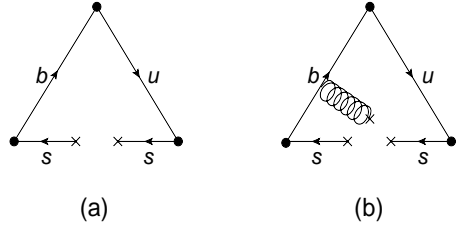}
\hspace{0.5cm}
\caption{The diagrams of the effective contributions of the condensate terms.} \label{F2}
\end{figure}

For calculation of the
non-perturbative contributions (condensate terms), we consider
the condensate terms of dimension $3, 4$ and $ 5$ related to
the contributions of the quark-quark, gluon-gluon and quark-gluon
condensate, respectively. They are more important than the other terms in
the OPE. In the 3PSR, when the light quark  is a spectator, the gluon-gluon condensate
contributions can be easily ignored \cite{PCol}.
On the other hand, the quark condensate contributions of the light quark which is a
non spectator, are zero after applying the double Borel
transformation with respect to the both variables $p^2$ and
$p'^2$, because only one variable appears in the denominator.
Therefore, only two important diagrams of dimension $3$, $4$ and $5$
remain from the non-perturbative part contributions.
The diagrams of these contributions corresponding to $C^{(3)}_{\alpha \beta \mu}$ and $C^{(5)}_{\alpha \beta \mu}$ are depicted in Fig. \ref{F2}.
After some calculations, the non-perturbative part of the correlation function are obtained as follows:
\begin{eqnarray}\label{eq217}
C^{(3)}_{V}+C^{(5)}_{V}&=&-\frac{2\kappa}{{(p^{2}-m_{b}^{2})}^{2}(p'^{2}-m_{u}^{2})},\nonumber\\
C^{(3)}_{0}+C^{(5)}_{0}&=&-\frac{4\kappa}{{(p^{2}-m_{b}^{2})}^{2}(p'^{2}-m_{u}^{2})},\nonumber\\
C^{(3)}_{1}+C^{(5)}_{1}&=&\frac{\kappa}{{(p^{2}-m_{b}^{2})}(p'^{2}-m_{u}^{2})}+\frac{\kappa[(m_{b}+m_{u})^{2}-q^{2}]}{{(p^{2}-m_{b}^{2})}^{2}(p'^{2}-m_{u}^{2})},\nonumber\\
C^{(3)}_{2}+C^{(5)}_{2}&=&-\frac{4\kappa}{{(p^{2}-m_{b}^{2})}^{2}(p'^{2}-m_{u}^{2})},
\end{eqnarray}
where $\kappa=\frac{(m_{s}^{2}-\frac{m_{0}^{2}}{2})}{48}\langle 0|s \bar s|0\rangle$,  $m^2_0=(0.8\pm 0.2) \mbox{GeV}^2$ \cite{PK}, and $\langle 0| \bar s s |0\rangle=(0.8\pm
0.2)\langle 0|\bar u u|0\rangle$, $\langle 0|\bar u u|0
\rangle=\langle 0|\bar d d|0\rangle=-(0.240\pm0.010~ \mbox{GeV})^3$ that we
choose the value of the condensates at a fixed renormalization scale
of about $1~ \mbox{GeV}$.

The next step is to apply the Borel transformations with respect to
the $p^2 (p^2\to M_1^2)$ and $p'^2 (p'^2\to M_2^2)$ on the
phenomenological as well as the perturbative and non-perturbative
parts of the correlation functions and equate these two
representations of the correlations. The following sum rules for the
form factors are derived:
\begin{eqnarray}\label{eq219}
V'(q^2)&=&\frac{(m_b+m_s)e^{m_{B_s}^2/M_1^2}e^{m_{K_2^*}^2/M_2^2}}{f_{B_s}m_{B_s}f_{K_2^*}m_{K_2^*}^2}
\left\{\frac{-1}{(2\pi)^{2}}\int^{s'_{0}}_{m_{s}^{2}}ds'\int^{s_{0}}_{s_{L}}ds\,\rho_{_V}(s,s',q^{2})e^{-s/M_1^2}e^{-s'/M_2^2}\right.\nonumber\\
&+&\left.\widetilde{B}\left[C_{V}^{(3)}+C_{V}^{(5)}\right]\right\},\nonumber\\
A'_n(q^2)&=&\frac{(m_b+m_s)e^{m_{B_s}^2/M_1^2}e^{m_{K_2^*}^2/M_2^2}}{f_{B_s}m_{B_s}f_{K_2^*}m_{K_2^*}^2}
\left\{\frac{-1}{(2\pi)^{2}}\int^{s'_{0}}_{m_{s}^{2}}ds'\int^{s_{0}}_{s_{L}}ds\,\rho_{n}(s,s',q^{2})e^{-s/M_1^2}e^{-s'/M_2^2}\right.\nonumber\\
&+&\left.\widetilde{B}\left[C_{n}^{(3)}+C_{n}^{(5)}\right]\right\},
\end{eqnarray}
where $n=0,...,2$, $s_{0}$ and $s'_{0}$ are the continuum thresholds in the initial and final channels, respectively. The lower limit in the integration over $s$ is:
$s_{L}=m_b^2+\frac{m_b^2}{m_b^2-q^2}s'$. Also $\widetilde{B}$ transformation is defined as follows:
\begin{eqnarray}\label{eq218}
\widetilde{B}\left[\frac{1}{{(p^2-m_b^2)}^{m}{(p'^2-m_u^2)}^{n}}\right]=
\frac{{(-1)}^{m+n}}{\Gamma(n)\Gamma(m)}\frac{e^{-m_b^2/M_1^2}e^{-m_u^2/M_2^2}}{{(M_1^2)}^{m-1}{(M_2^2)}^{n-1}},
\end{eqnarray}
where $M_{1}^{2}$ and $M_{2}^{2}$ are Borel mass parameters.

We would like to provide the same results for the $B\to a_2
\ell \nu$, and $B\to f_2 \ell \nu$ decays. With a little bit of
change in the above expressions such as $s\leftrightarrow d(u)$ and $m_{K^*_2}\leftrightarrow m_{a_2}(m_{f_2})$, we can easily find  similar results
in Eq. (\ref{eq219}) for the form factors of the new transitions.

\section{Numerical analysis}

In this section, we numerically analyze the sum rules for the form
factors $V(q^{2})$, $A_0(q^{2})$, $A_1(q^{2})$ and $A_2(q^{2})$ as well as
branching ratio values of the transitions $B (B_s)\to T$, where $T$ can be one
of the tensor mesons $K_{2}^{*},~ a_2$, or $f_2$. The
values of the meson masses and leptonic decay constants are chosen as presented
in Table \ref{T31}.
\begin{table}[th]
\caption{The values of the meson masses  \cite{PDG} and decay
constants \cite{Wang} in $\mbox{GeV}$.}\label{T31}
\begin{ruledtabular}
\begin{tabular}{cccccccccc}
${\rm meson}$          & $B_s$   & $B$  & $K_2^*$ & $a_2$ & $f_2$\\
\hline
${\rm Mass}$           & $5.366$ & $5.279$ & $1.425$     & $1.318$   & $1.275$\\
${\rm Decay~  Constant}$ & $0.230$ & $0.190$ & $0.118$     & $0.107$
& $0.102$
\end{tabular}
\end{ruledtabular}
\end{table}
Also $m_{b}=4.820~\mbox{GeV}$, $m_{s}=0.150~\mbox{GeV}$ \cite{MQH},
$m_{\tau}=1.776~\mbox{GeV}$, and $m_{\mu}=0.105~\mbox{GeV}$
\cite{PDG}.

From the 3PSR, it is clear that the form factors also contain the
continuum thresholds $s_{0}$ and $s'_{0}$ and the Borel parameters
$M_{1}^{2}$ and $M_{2}^{2}$ as the main input. These are not
physical quantities, hence the form factors, should be independent
of these parameters. The continuum thresholds, $s_{0}$ and $s_{0}'$
are not completely arbitrary, but these are in correlation with the
energy of the first exited state with the same quantum numbers as
the considered interpolating currents. The value of the continuum
threshold $s^{B(B_s)}_0=35~\mbox{GeV}^2$ \cite{Shifman} calculated
from the 3PSR. The values of the continuum threshold $s'_0$ for the
tensor mesons $K_2^*$, $a_2$ and $f_2$ are taken to be
$s^{K_2^*}_0=3.13 ~\mbox{GeV}^2$, $s^{a_2}_0=2.70 ~\mbox{GeV}^2$ and
$s^{f_2}_0=2.53 ~\mbox{GeV}^2$, respectively \cite{Cheng}.

\begin{figure}
\includegraphics[width=6cm,height=5cm]{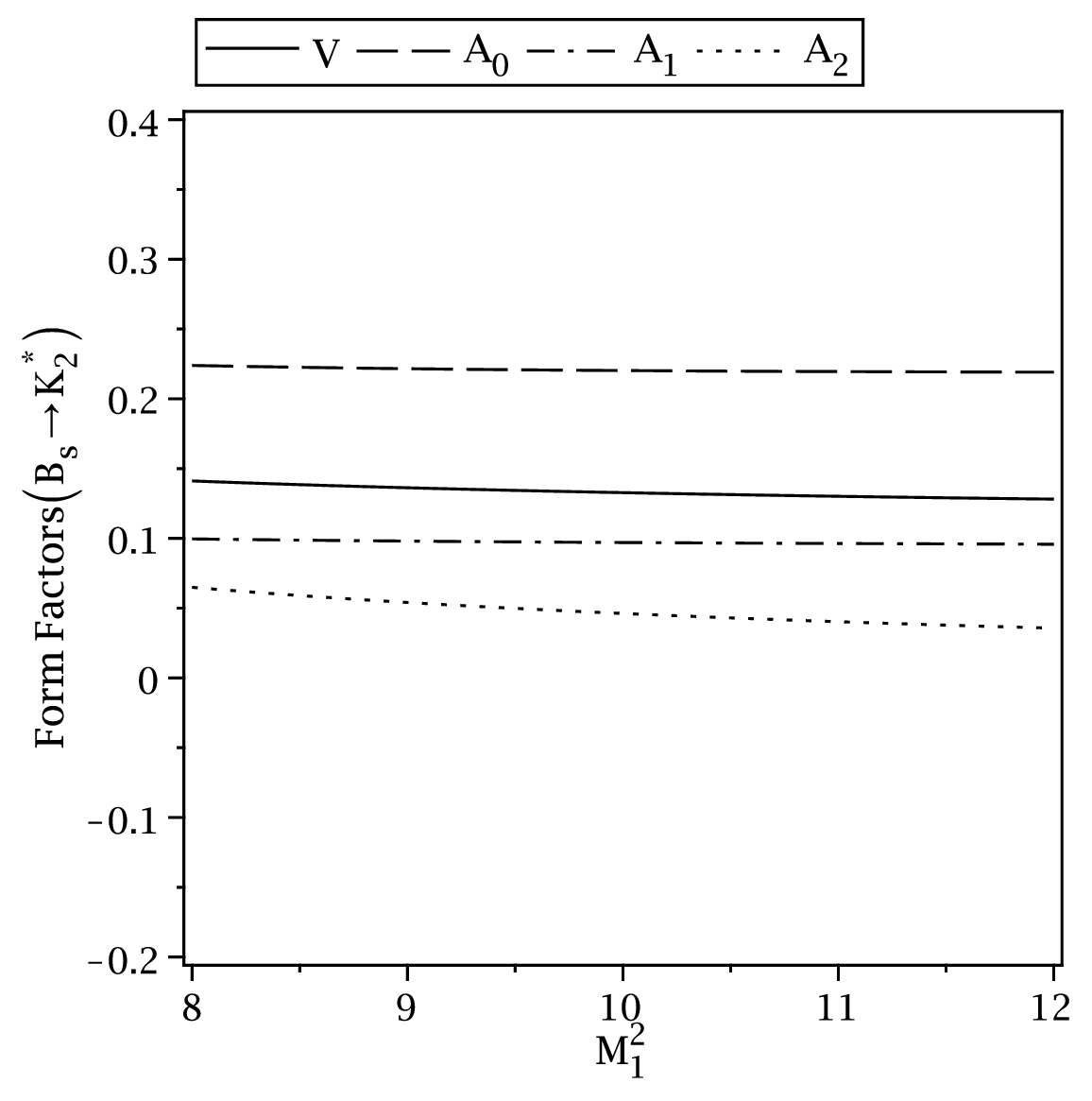}
\includegraphics[width=6cm,height=5cm]{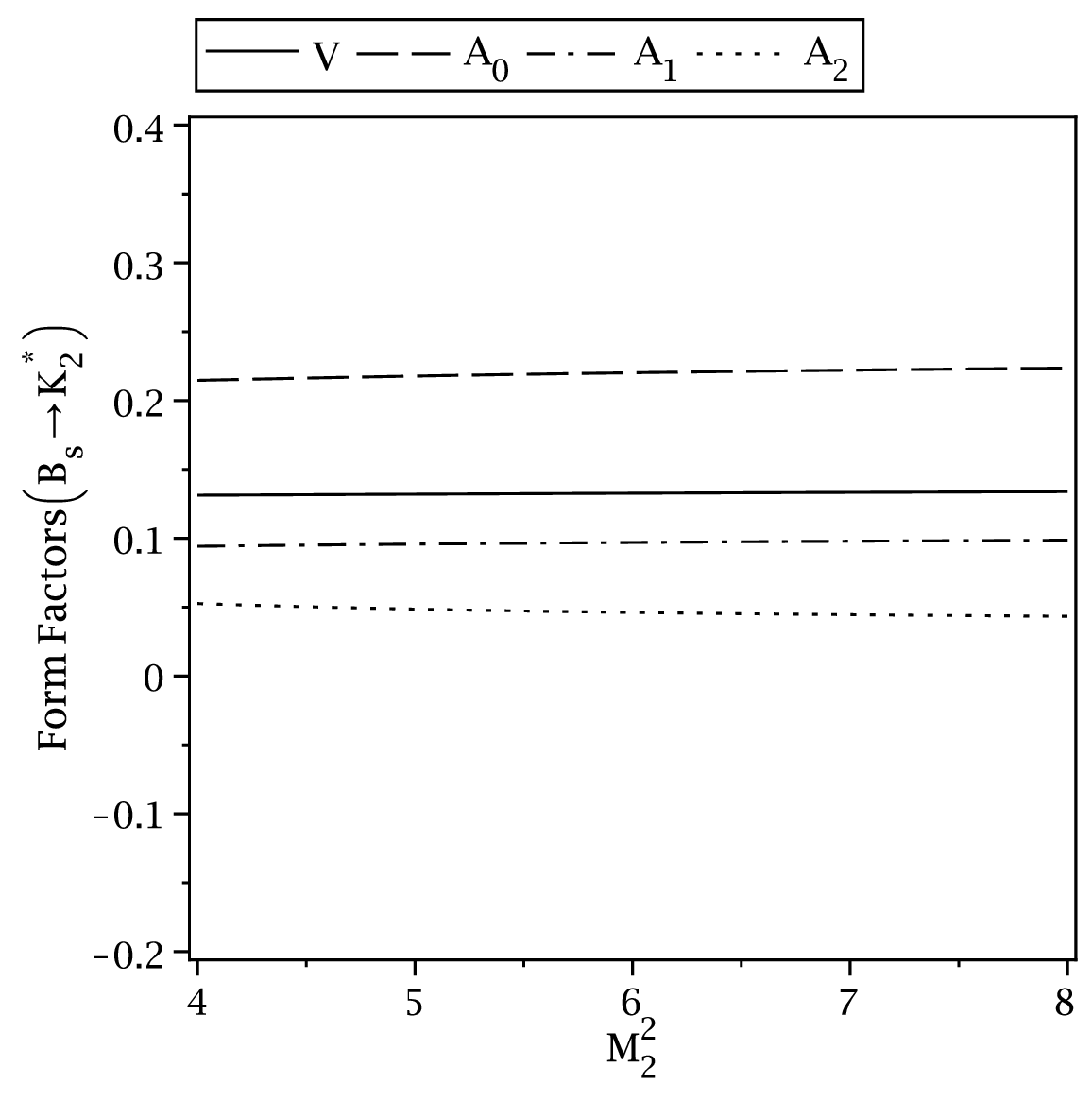}
\caption{The form factors of $B_s\to K_2^*$ on $M_1^{2}$ and
$M_2^{2}$.} \label{F31}
\end{figure}

We search for the intervals of the Borel parameters so that our
results are almost insensitive to their variations. One more
condition for the intervals of these parameters is the fact that the
aforementioned intervals must suppress the higher states, continuum
and contributions of the highest-order operators. In other words,
the sum rules for the form factors must converge.  As a result, we
get $8~\mbox{GeV}^{2}\leq M_{1}^{2} \leq 12~\mbox{GeV}^2$ and
$4~\mbox{GeV}^2\leq M_{2}^{2}\leq 8~\mbox{GeV}^2$. To show how the
form factors depend on the Borel mass parameters, as examples, we
depict the variations of the form factors $V$, $A_{0}$, $A_{1}$ and
$A_{2}$ for $B_{s}\rightarrow K_{2}^{*}\ell\nu$ at $q^{2} = 0$ with
respect to the variations of the $M_1^2$ and $M_2^2$ parameters in
their working regions in Fig. \ref{F31}. From these figures, it
revealed that the form factors weakly depend on these parameters in
their working regions.

\begin{figure}
\includegraphics[width=5cm,height=4cm]{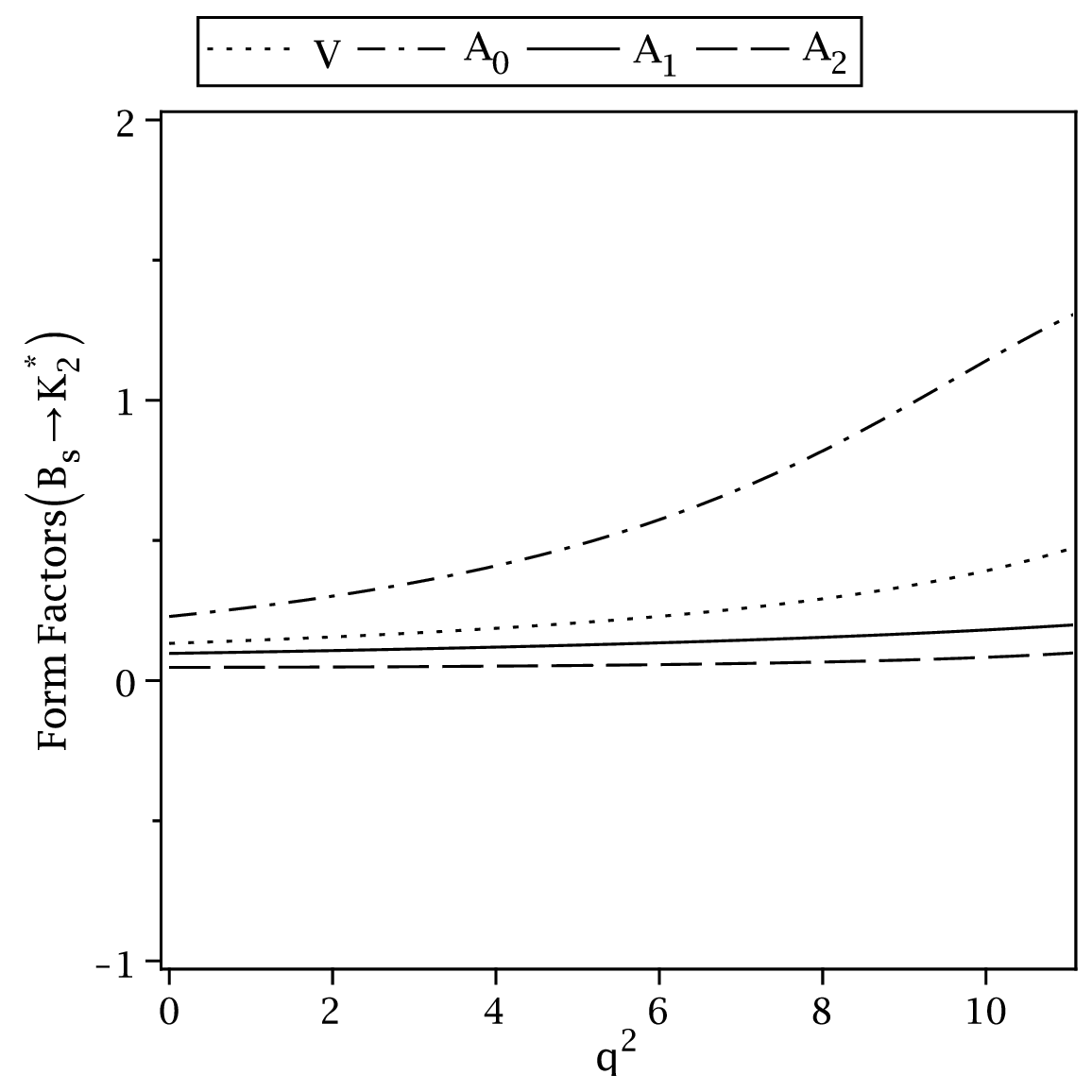}
\includegraphics[width=5cm,height=4cm]{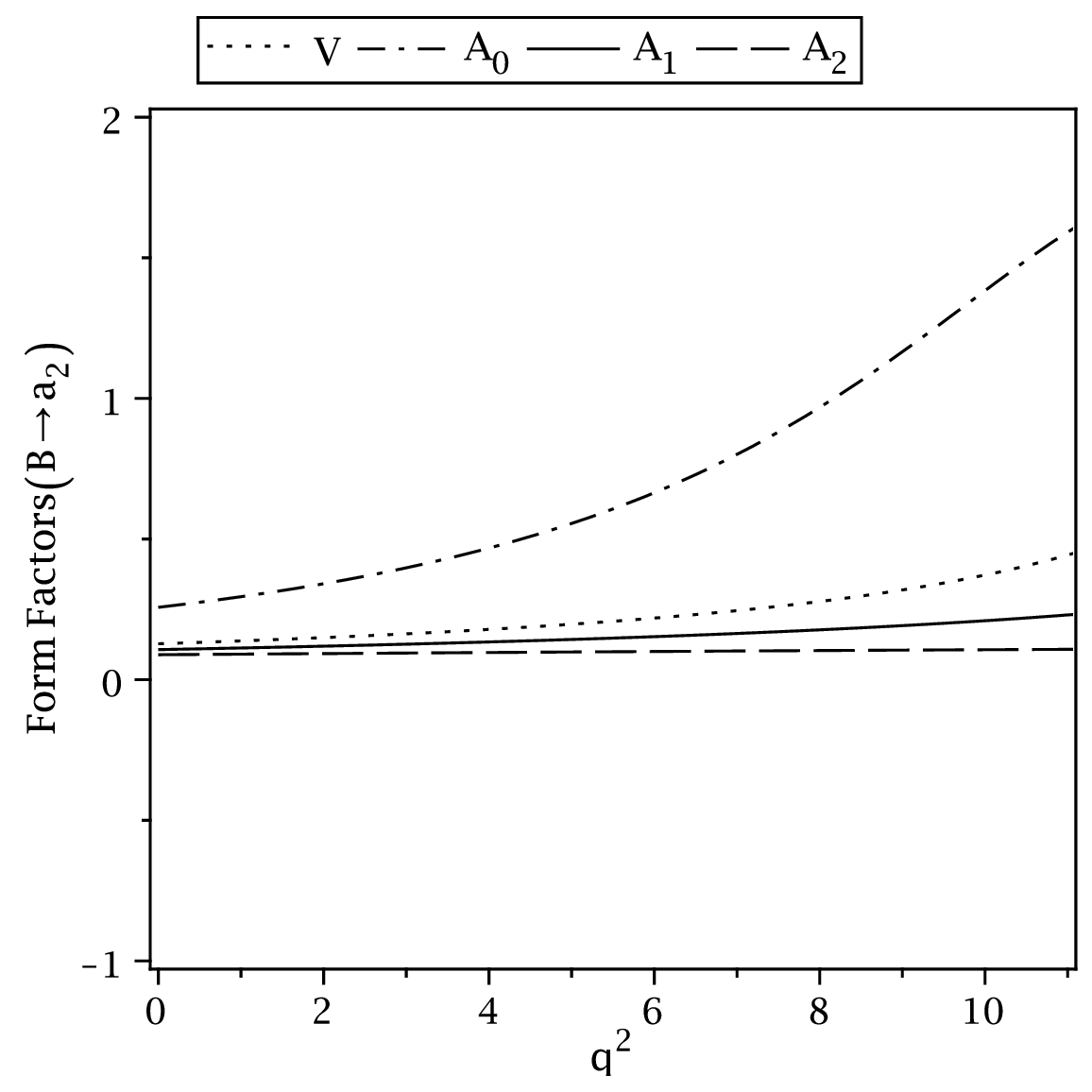}
\includegraphics[width=5cm,height=4cm]{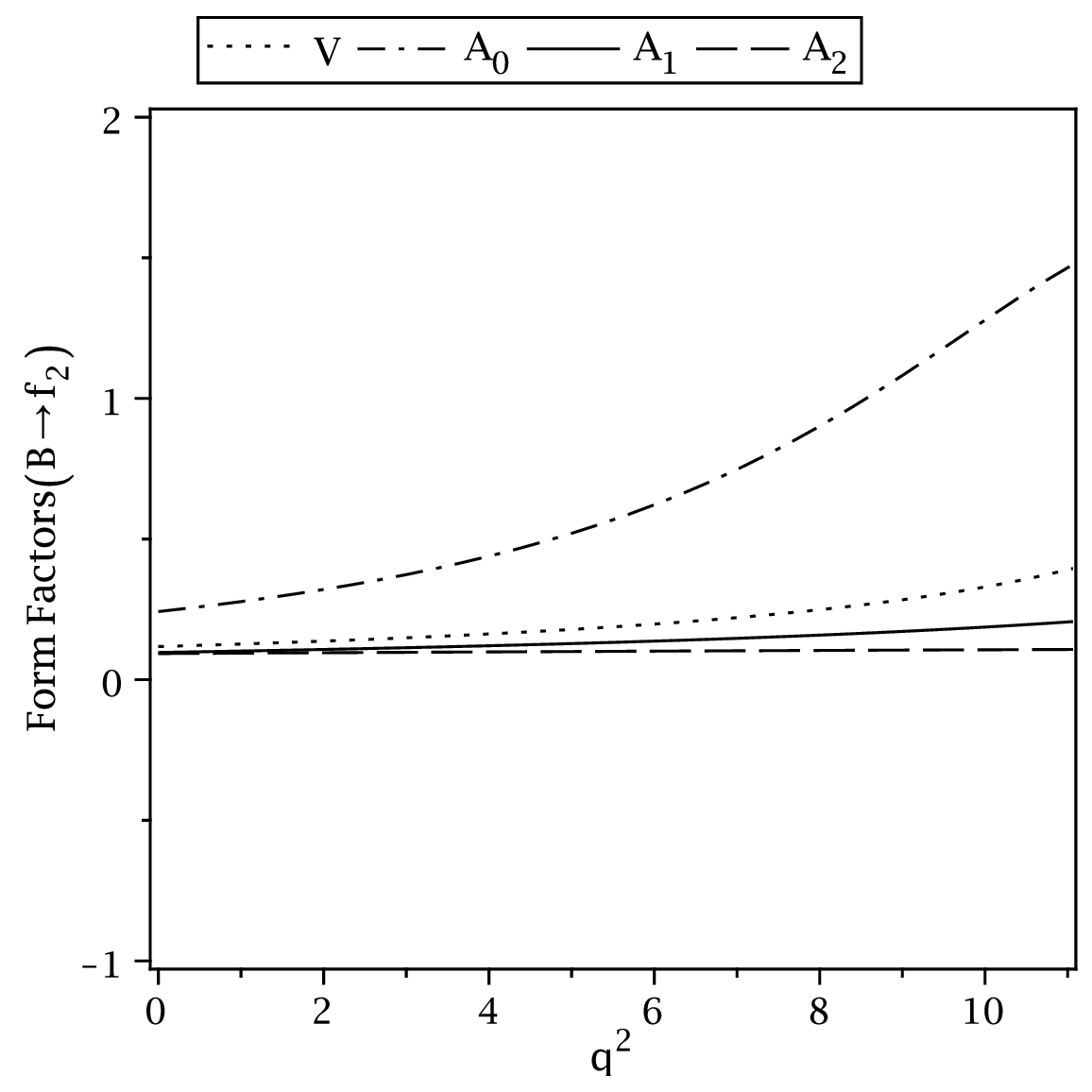}
\caption{The SR predictions for the form factors of the $B(B_s)\to T
\ell \nu$ transitions on $q^2$.} \label{F32}
\end{figure}

The sum rules for the form factors are truncated at about $0\leq
q^2\leq 11~ {\rm GeV}^2$. The dependence of the form factors $V$,
$A_{0}$, $A_{1}$ and $A_{2}$ on $q^{2}$ for $B\to T$ transitions are
shown in Fig. \ref{F32}.
However, it is necessary to obtain the behavior of the form factors
with respect to $q^2$ in the full physical region, $0\leq q^{2}
\leq(m_{B(B_{s})}-m_{T})^{2}$, in order to calculate the decay width
of the $B \to T$ transitions. So to extend our results, we look for
a parametrization of the form factors in such a way that in the
region $0\leq q^{2} \leq(m_{B(B_{s})}-m_{T})^{2}$, this
parametrization coincides with the sum rules predictions. Our
numerical calculations show that the sufficient parametrization of
the form factors with respect to $q^2$ is as follows:
\begin{equation}
f(q^{2})=\frac{f(0)}{1-a(\frac{q^{2}}{m_{B(B_{s})}^{2}})+b(\frac{q^{2}}{m_{B(B_s)}^2})^{2}}.
\end{equation}
The values of the parameters $f(0)$, $a$, and $b$ for the transition
form factors of the $B\to T$ are given in the Table \ref{T32}.
\begin{table}[th]
\caption{Parameter values appearing in the fit functions of the
$B\to T \ell \nu$ decays.}\label{T32}
\begin{ruledtabular}
\begin{tabular}{ccccccccc}
${\rm Form~ Factor}$ & $f(0)$ & $a$ & $b$ & ${\rm Form~ Factor}$ & $f(0)$ & $a$ & $b$\\
\hline
$V^{B_s\to K_2^*}$ & $0.13$ & $2.19$ & $0.83$    & $A_0^{B_s\to K_2^*}$ & $0.23$ & $3.77$ & $4.21$\\
$A_1^{B_s\to K_2^*}$ & $0.10$ & $1.36$ & $0.09$ & $A_2^{B_s\to K_2^*}$ & $0.05$ & $0.21$ & $-2.99$\\
$V^{B\to a_2}$ & $0.13$ & $2.10$ & $0.75$        & $A_0^{B\to a_2}$ & $0.26$ & $3.71$ & $4.03$\\
$A_1^{B\to a_2}$ & $0.11$ & $1.45$ & $0.23$     & $A_2^{B\to a_2}$ & $0.09$ & $0.63$ & $0.46$\\
$V^{B\to f_2}$ & $0.12$ & $2.01$ & $0.60$        & $A_0^{B\to f_2}$ & $0.24$ & $3.70$ & $4.02$\\
$A_1^{B\to f_2}$ & $0.10$ & $1.40$ & $0.16$     & $A_2^{B\to f_2}$ & $0.09$ & $0.46$ & $0.29$\\
\end{tabular}
\end{ruledtabular}
\end{table}

In Table \ref{T35}, our results for the form factors of $B\to T \ell
\nu$ decays in $q^2=0$ is compared with those of other approaches
such as the LCSR, the PQCD, the LEET, and the ISGW II model. Our
results are in good agreement  with those of the LCSR, PQCD and LEET
in all cases.
\begin{table}[th]
\caption{Comparison of the form factor values of the $B\to T \ell
\nu$ decays in $q^2=0$ in different approaches.}\label{T35}
\begin{ruledtabular}
\begin{tabular}{ccccccccc}
${\rm Form~ Factor}$ & {\rm This Work} & LCSR\cite{Yang} & PQCD\cite{Wang} & LEET\cite{Charles,Ebert,Datta} & ISGW II\cite{Scora}  \\
\hline
$V^{B_s\to K_2^*}$   &$0.13\pm0.03$&     $0.15\pm0.02 $        &${0.18}^{+0.05}_{-0.04} $   &   $-$            & $-$       \\
$A_0^{B_s\to K_2^*}$ &$0.23\pm0.06$&     $0.22\pm0.04 $        &${0.15}^{+0.04}_{-0.03} $   &   $-$            & $-$       \\
$A_1^{B_s\to K_2^*}$ &$0.10\pm0.02$&     $0.12\pm0.02 $        &${0.11}^{+0.03}_{-0.02} $   &   $-$            & $-$       \\
$A_2^{B_s\to K_2^*}$ &$0.05\pm0.01$&     $0.05\pm0.02 $        &${0.07}^{+0.02}_{-0.02} $   &   $-$            & $-$       \\
$V^{B\to a_2}$       &$0.13\pm0.03$&     $0.18\pm0.02 $        &${0.18}^{+0.05}_{-0.04} $   &   $0.18\pm 0.03$ & $0.32$    \\
$A_0^{B\to a_2}$     &$0.26\pm0.07$&     $0.21\pm0.04 $        &${0.18}^{+0.06}_{-0.04} $   &   $0.14\pm 0.02$ & $0.20$    \\
$A_1^{B\to a_2}$     &$0.11\pm0.04$&     $0.14\pm0.02 $        &${0.11}^{+0.03}_{-0.03} $   &   $0.13\pm 0.02$ & $0.16$    \\
$A_2^{B\to a_2}$     &$0.09\pm0.02$&     $0.09\pm0.02 $        &${0.06}^{+0.02}_{-0.01} $   &   $0.13\pm 0.02$ & $0.14$    \\
$V^{B\to f_2}$       &$0.12\pm0.04$&     $0.18\pm0.02 $        &${0.12}^{+0.03}_{-0.03} $   &   $0.18\pm 0.02$ & $0.32$    \\
$A_0^{B\to f_2}$     &$0.24\pm0.06$&     $0.20\pm0.04 $        &${0.13}^{+0.04}_{-0.03} $   &   $0.13\pm 0.02$ & $0.20$    \\
$A_1^{B\to f_2}$     &$0.10\pm0.02$&     $0.14\pm0.02 $        &${0.08}^{+0.02}_{-0.02} $   &   $0.12\pm 0.02$ & $0.16$    \\
$A_2^{B\to f_2}$     &$0.09\pm0.02$&     $0.10\pm0.02 $
&${0.04}^{+0.01}_{-0.01} $   &   $0.13\pm 0.02$ & $0.14$
\end{tabular}
\end{ruledtabular}
\end{table}

At the end of this section, we would like to present the
differential decay widths of the process under consideration. Using
the parametrization of these transitions in terms of the form
factors, the differential decay width for $B \rightarrow T \ell \nu$
transition is obtained as:
\begin{equation}\label{eq221}
\frac{d\Gamma(B\to T \ell \nu)}{dq^2}=\frac{\mid
G_{F}V_{ub}\mid^{2}\sqrt{\lambda(m_{B}^{2},m_{T}^{2},q^2)}}{256~
m_{B}^{3}~\pi^{3}q^{2}}(1-\frac{m_{\ell}^{2}}{q^{2}})^{2}(X_{L}+X_{+}+X_{-}),
\end{equation}
$m_{\ell}$ represents the mess of the charged lepton. The other
parameters are defined as
\begin{eqnarray*}\label{eq222}
X_{L}&=&\frac{1}{9}\frac{\lambda}{m_{T}^2~ m_{B}^{2}}[(2q^2+m_{\ell}^{2})h_{0}^{2}(q^{2})+3\lambda m_{\ell}^{2} A_{0}^{2}(q^{2})],\nonumber\\
X_{\pm}&=&\frac{2q^2}{3}(2 q^2 + m_{\ell}^{2})\frac{\lambda}{8m_{T}^2~ m_{B}^{2}}[(m_{B}+m_{T})A_{1}(q^{2})\mp\frac{\sqrt{\lambda}}{m_{B}+m_{T}}V(q^2)]^{2},\nonumber\\
h_{0}(q^{2})&=&\frac{1}{2 m_{T}}[(m_{B}^2 - m_{T}^2 -
q^{2})(m_{B}+m_{T})A_{1}(q^{2})-\frac{\lambda}{m_{B}+m_{T}}A_{2}(q^{2})].
\end{eqnarray*}
\begin{figure}
\includegraphics[width=6cm,height=5cm]{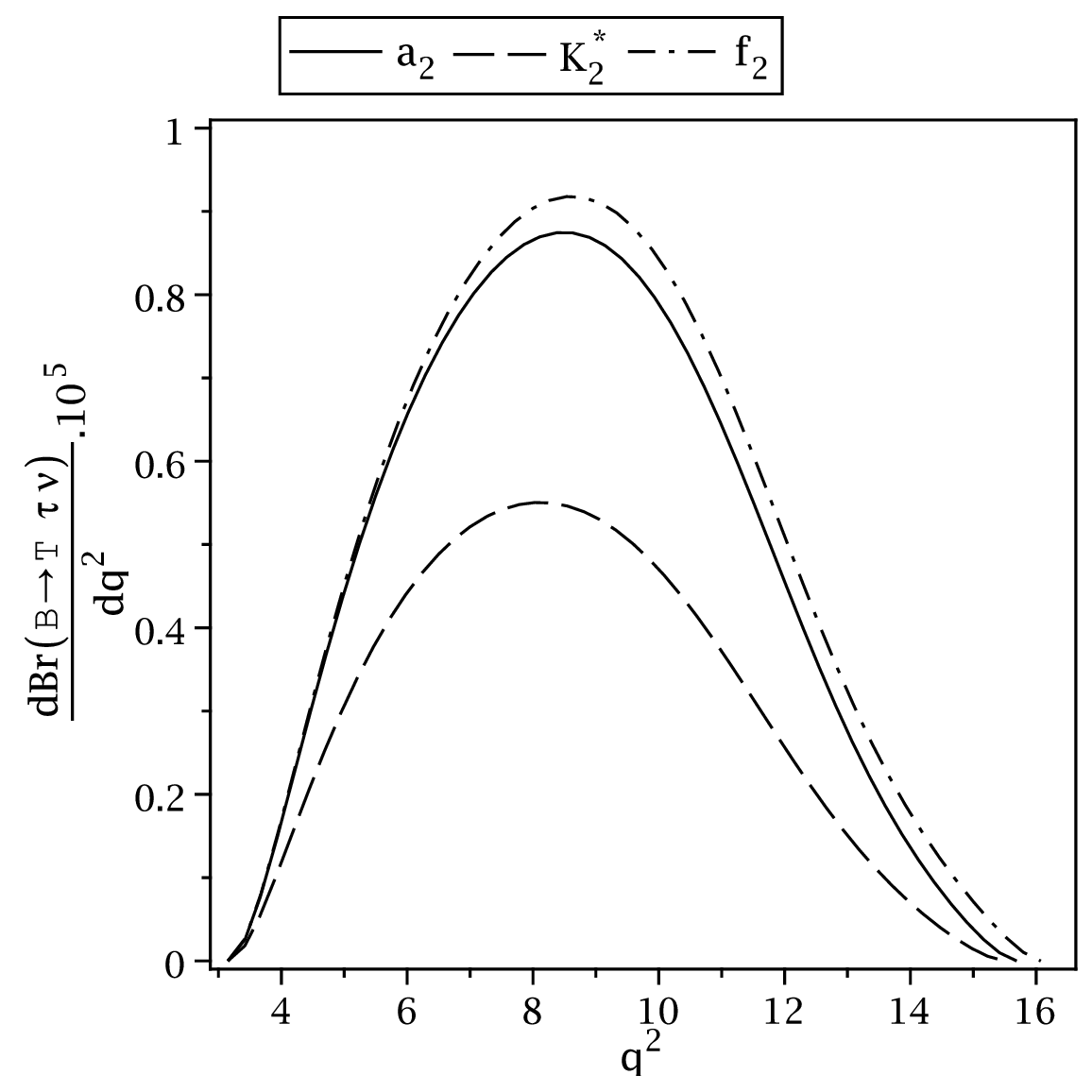}
\includegraphics[width=6cm,height=5cm]{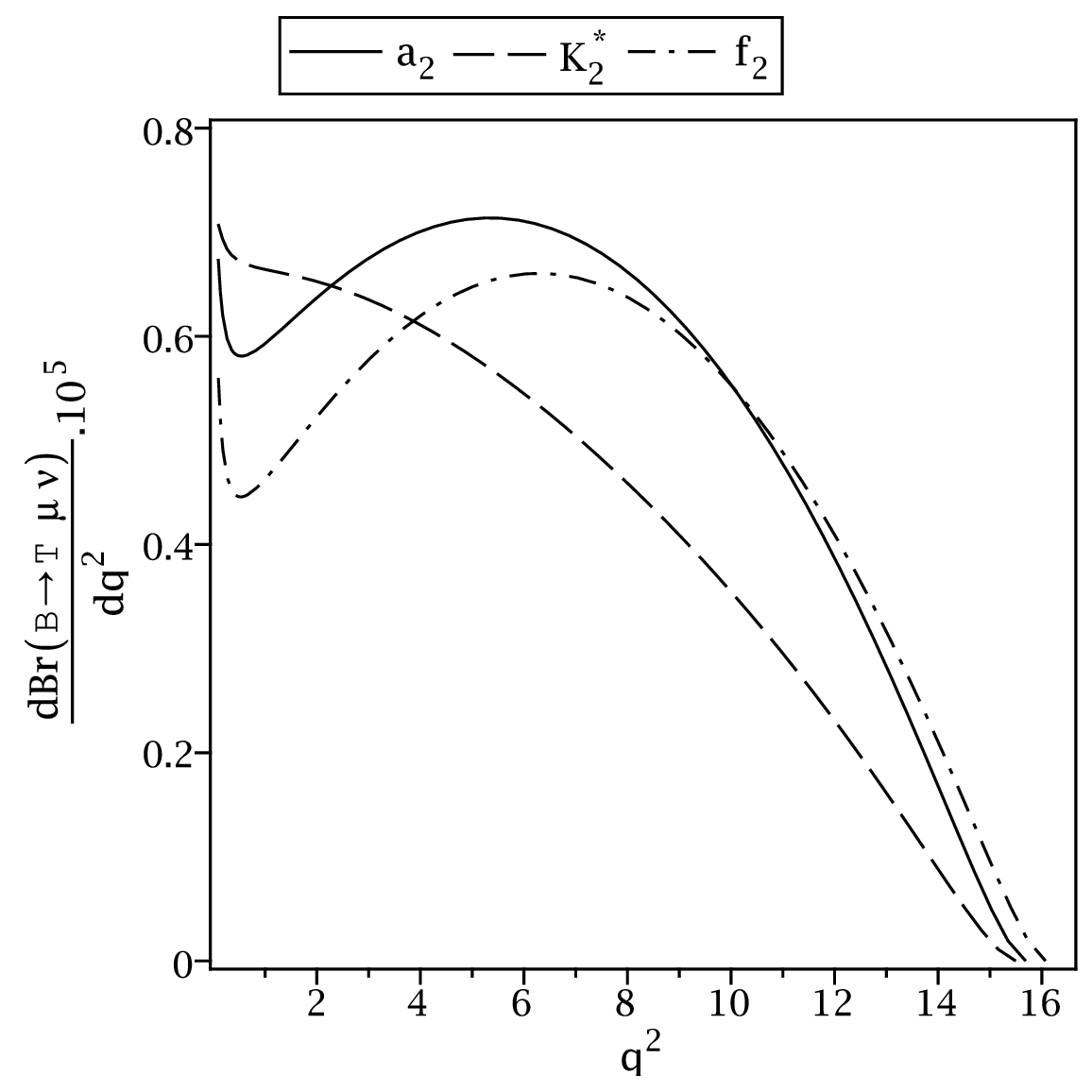}
\caption{The differential branching ratios of the semileptonic $B\to
T \ell \nu$ decays on $q^2$.} \label{F33}
\end{figure}
Integrating Eq. (\ref{eq221}) over $q^{2}$ in the whole physical
region, and using $V_{ub}=(3.89\pm0.44)\times10^{-3}$ \cite{PDG},
the branching ratios of the $B\rightarrow T\ell\nu$ are obtained.
The differential branching ratios of the $B\to T \ell \nu$ decays on
$q^{2}$ are shown in Fig. \ref{F33}. The branching ratio values
of these decays  are also obtained as
presented in Table \ref{T36}. Furthermore, this table contains the results
estimated via the PQCD. Considering the uncertainties, our
estimations for the  branching ratio values of the $B\to T \ell \nu$
decays are  in consistent agreement with those of the PQCD.
\begin{table}[th]
\caption{Comparison of the branching ratio values of the $B\to T
\ell \nu$ decays with those of the PQCD (in
units of $10^{-4}$).}\label{T36}
\begin{ruledtabular}
\begin{tabular}{cccccc}
&{\rm This Work}& PQCD\cite{Wang}\\
\hline
${\rm Br}(B\to a_2 \mu \nu)$&$0.82\pm 0.25$&${1.16}^{+0.81}_{-0.57}$\\
${\rm Br}(B_s\to K_2^* \mu \nu)$&$0.65\pm0.20$&${0.73}^{+0.48}_{-0.33}$\\
${\rm Br}(B\to f_2 \mu \nu)$&$0.77\pm0.23$&${0.69}^{+0.48}_{-0.34}$\\
${\rm Br}(B\to a_2 \tau \nu)$&$0.51\pm0.17$&${0.41}^{+0.29}_{-0.20}$\\
${\rm Br}(B_s\to K_2^* \tau \nu)$&$0.35\pm0.11$&${0.25}^{+0.17}_{-0.12}$\\
${\rm Br}(B\to f_2 \tau \nu)$&$0.53\pm0.18$&${0.25}^{+0.18}_{-0.13}$\\
\end{tabular}
\end{ruledtabular}
\end{table}

In summary, we considered the $B_{s}(B) \to K_2^*(a_2, f_2) \ell\nu
$ channels and computed the relevant form factors considering the
contribution of the quark condensate corrections. Our results are in good agreement  with those of the LCSR, PQCD and LEET in
all cases. We also evaluated
the total decays widthes and the branching ratios of these decays.
Our branching ratio values of these decays are in consistent
agreement with those of the PQCD.

\section*{Acknowledgments}
Partial support of Isfahan university of  technology research council is appreciated.
\clearpage

\appendix
\begin{center}
{\Large \textbf{Appendix}}
\end{center}
In this appendix,  the explicit expressions of the coefficients
$\lambda(s, s', q^2)$, $B_l~ (l=1, 2)$, $D_j~ (j=1,...,4)$, and
$E_r~(r=1,...,6)$, are given.
\begin{eqnarray*}
\lambda(s,s',q^2)&=&s^2+ s'^2+ (q^{2})^2- 2sq^{2}- 2s'q^{2}- 2ss',\nonumber\\
B_1 &=& \frac{I_0}{\lambda(s,s',q^2)} \left [2 s' \Delta -\Delta' u\right],\nonumber \\
B_2 &=& \frac{I_0}{\lambda(s,s',q^2)} \left [2 s \Delta' -\Delta u\right],\nonumber \\
D_1 &=& -\frac{I_0}{2\lambda(s,s',q^2)}[4ss'm_s^2-s\Delta'^2-s'\Delta^2-u^2m_s^2+u\Delta\Delta'],\nonumber \\
D_2 &=& -\frac{I_0}{\lambda^2(s,s',q^2)}[8ss'^2m_s^2-2ss'\Delta'-6s'^2\Delta^2-2u^2s'm_s^2
+6s'u\Delta\Delta'-u^2\Delta'^2],\nonumber \\
D_3 &=& \frac{I_0}{\lambda^2(s,s',q^2)} \left [4 s s' u m_s^2+4 s s' \Delta \Delta'-3 s u \Delta'^2-3u \Delta^2 s'-u^3 m_3^2+2 u^2
\Delta\Delta'\right],\nonumber \\
D_4 &=&\frac{I_0}{\lambda^2(s,s',q^2)}[-6s'u\Delta\Delta'+6s^2\Delta'^2-8s^2s'm_s^2+2u^2s~m_s^2 +u^2\Delta^2+2ss'\Delta^2],\nonumber\\
E_1 &=& \frac{I_0}{2\lambda^2(s,s',q^2)}\left[ 8 s'^2 m_s^2\Delta s-2 s'm_s^2\Delta u^2-4 u m_s^2 \Delta' s s'
+u^3 m_s^2\Delta' -2 s'^2 \Delta^3\right.\nonumber\\ &+&\left.3 s' u \Delta^2 \Delta' -2\Delta'^2 \Delta s s'-\Delta'^2 \Delta u^2+u s\Delta^3\right],\nonumber\\
E_2 &=& \frac{I_0}{2\lambda^2(s,s',q^2)}\left[ 8 s^2 m_s^2 \Delta' s'-2 s^2 \Delta'^3- 4 u m_s^2 \Delta s s'-2 \Delta^2 \Delta' s s'+3 u s \Delta'^2 \Delta  \right.\nonumber\\
&-&\left. 2 s m_s^2 \Delta' u^2+s' u \Delta^3+u ^3  m_s^2 \Delta-\Delta^2 \Delta' u^2\right], \nonumber\\
E_3 &=&-\frac{I_0}{{\lambda^3(s,s',q^2)}} \left[ 48 s m_s^2 \Delta s'^3 -24 s s'^2 u m_s^2 \Delta'-12 s s'^2 \Delta'^2 \Delta
+6 s u \Delta'^3 s'-20 s'^3\Delta^3 \right. \nonumber\\ &+&\left.
30 s'^2 u \Delta^2 \Delta'-12 s'^2 m_s^2 \Delta u^2 -12 s' \Delta'^2 \Delta u^2+6 s' u^3 m_s^2 \Delta'+ u^3 \Delta'^3\right], \nonumber\\
E_4 &=& -\frac{I_0}{\lambda^3(s,s',q^2)}\left[ 16 s^2 m_s^2 \Delta' s'^2- 4 s^2 \Delta'^3 s' -12 s s'^2 \Delta^2 \Delta'
-24 s s'^2 u m_s^2 \Delta + 3 u^3 \Delta'^2 \Delta \right. \nonumber\\  &+&\left.
18 s u \Delta'^2 \Delta  s' -4 s \Delta'^3 u^2+10 s'^2 u \Delta^3 +6 s' u^3 m_s^2 \Delta-12 s' \Delta^2 \Delta' u^2
- 2 m_s^2 \Delta' u^4\right.\nonumber\\&+&\left.4 s s' u^2 m_s^2 \Delta'\right],\nonumber\\
E_{5} &=& -\frac {I_0}{\lambda^3(s,s',q^2)}\left[ 16 s^2 m_s^2 \Delta s'^2-24 s^2 s' u m_s^2 \Delta'
-12 s^2 s' \Delta'^2 \Delta +10 u s^2 \Delta'^3 - 4 s s'^2 \Delta^3 \right.\nonumber\\&+&\left. 4 s s' u^2 m_s^2 \Delta+18 s  u \Delta^2 \Delta' s'
+6 s u^3 m_s^2 \Delta' -12 s \Delta^2 \Delta u^2- 4 s' \Delta^3 u^2- 2 m_s^2 \Delta u^4\right. \nonumber\\&+&\left.3 u^3 \Delta^2 \Delta' \right],\nonumber\\
E_{6} &=& -\frac{I_0}{\lambda^3(s,s',q^2)}\left[ 48 s^3 m_s^2\Delta' s'-20 s^3 \Delta'^3- 12 s^2 \Delta^2 \Delta' s'
-24 s^2 s' u m_s^2 \Delta - 12 s^2 m_s^2 \Delta' u^2\right.\nonumber\\ &+&\left. 30 u s^2 \Delta'^2 \Delta +6 s u \Delta^3 s'
-12 s \Delta^2 \Delta' u^2+6 s u^3 m_s^2 \Delta+ u^3 \Delta^3\right],\nonumber\\
\Delta &=& s+m_s^2-m_b^2,~~~~~~~ \Delta'=s'+m_s^2-m_u^2,~~~~~~~ u=s+s'-q^2.
\end{eqnarray*}


\begin{thebibliography}{99}

\bibitem{Aubert1}
B. Aubert  et al., BABAR Collaboration, Phys. Rev. D 79, 052005
(2009)

\bibitem{Aubert2}
B. Aubert  et al.,  BABAR Collaboration, Phys. Rev. Lett. 101,
161801 (2008).

\bibitem{Aubert3}
B. Aubert  et al.,  BABAR Collaboration,  Phys. Rev. D 78, 092008
(2008).

\bibitem{Barberio}
E. Barberio  et al., Heavy Flavor Averaging Group (HFAG), arXiv:
0808.1297.

\bibitem{Wang}
W. Wang, Phys. Rev. D 83, 014008 (2011).

\bibitem{Dombrowski}
S. V. Dombrowski, Nucl. Phys. Proc. Suppl. 56, 125 (1997).

\bibitem{PDG1}
C. Amsler et al., Particle Data Group, Phys. Lett. B 667, 1 (2008).

\bibitem{Li}
D. M. Li, H. Yu, and Q. X. Shen, J. Phys. G 27, 807 (2001).

\bibitem{Cheng}
H. Y. Cheng, Y. Koike, and K. C. Yang, Phys. Rev. D 82, 054019
(2010).

\bibitem{Ball1}
P. Ball, Phys. Rev. D 48, 3190 (1993).

\bibitem{ColFa}
P. Colangelo, F. De Fazio, P. Santorelli and E. Scrimieri, Phys.
Rev. D 53, 3672 (1996); Phys. Lett. B 395, 339 (1997).

\bibitem{CoDom}
P. Colangelo, C. A. Dominguez, G. Nardulli and N. Paver, Phys. Lett.
B 317, 183 (1993).

\bibitem{Yang2}
M. Z. Yang, Phys. Rev. D 73, 034027 (2006).

\bibitem{Khosravi1}
N. Gharamany and R. Khosravi, Phys.  Rev,  D 80, 016009 (2009).

\bibitem{Khosravi2}
R. Khosravi, Eur. Phys. J. C 75: 220-229, (2015).

\bibitem{Ball3}
P. Ball, arXiv: hep-ph/9308244.

\bibitem{CLEO1}
T. E. Coan et al., CLEO Collaboration, Phys. Rev. Lett. 84, 5283
(2000)

\bibitem{CLEO2}
S. Ahmed et al., CLEO Collaboration, arXiv: hep-ex/9908022.

\bibitem{ALEPH}
R. Barate et al., ALEPH Collaboration, Phys. Lett. B 429, 169
(1998).

\bibitem{Abada}
A. Abada et al., ELC Collaboration, Nucl. Phys. B 416, 675 (1994).

\bibitem{Allton}
C. R. Allton et al. APE Collaboration, Phys. Lett. B 345, 513
(1995).

\bibitem{UKQCD}
D. R. Burford et al., UKQCD Collaboration, Nucl. Phys. B 447, 425
(1995).

\bibitem{PCol}
P. Colangelo and A. Khodjamirian, arXiv: hep-ph/0010175; A. V.
Radyushkin, arXiv: hep-ph/0101227.

\bibitem{Ali}
A. Ali, V. M. Braun, H. Simma, Z. Phys. C 63, 437 (1994).

\bibitem{Ball4}
P. Ball, V. M. Braun, Phys. Rev. D 55, 5561, (1997).

\bibitem{Ball5}
P. Ball, TU-Munchen Report TUM-T31-43/93.

\bibitem{Yang}
K. C. Yang, Phys. Lett. B 695, 444 (2011).

%\bibitem{Cheng1}
%H. Y. Cheng, C. K. Chua and C. W. Hwang,  Phys. Rev.  D  69, 074025
%(2004).
%
%\bibitem{Cheng2}
%H. Y. Cheng and C. K. Chua,  Phys.  Rev.  D  69, 094007 (2004)
%[Erratum-ibid.  D  81, 059901 (2010)].
%
%\bibitem{Cheng3}
%H. Y. Cheng and C. K. Chua,  Phys. Rev.  D 81, 114006 (2010).
%
%\bibitem{Sharma}
%N. Sharma and R. C. Verma, Phys. Rev. D 82, 094014 (2010).

\bibitem{Charles}
J. Charles, A. L. Yaouanc, L. Oliver, O. Pene and J. C. Raynal,
Phys. Lett. B 451, 187 (1999).

\bibitem{Ebert}
D. Ebert, R. N. Faustov and V. O. Galkin, Phys. Rev. D 64, 094022
(2001).

\bibitem{Datta}
A. Datta, Y. Gao, A. V. Gritsan, D. London, M. Nagashima and A.
Szynkman, Phya. Rev. D 77, 114025 (2008).

\bibitem{Scora}
D. Scora and N. Isgur, Phys. Rev. D 52, 2783 (1995).

\bibitem{Ball}
P. Ball and R. Zwicky, Phys. Rev. D 71, 014029 (2005).

\bibitem{PK}
P. Colangelo and A. Khodjamirian, in \emph{At the Frontier of
Particle Physics/Handbook of QCD}, edited by M. Shifman (World
Scientific, Singapore, 2001), Vol. III, p. 1495.

\bibitem{MQH}
M. Q. Huang,  Phys. Rev. C 69, 114015 (2004).

\bibitem{PDG}
J. Beringer et al., Particle Data Group, Phys. Rev. D 86, 010001
(2012).

\bibitem{Shifman}
M. A. Shifman, A. I. Vainshtein and V. I. Zakharov, Nucl, Phys, B
147, 385, (1979)

%\bibitem{Navara2}
%Z. Guo, S. Narison, J. M. Richard and Q. Zhao, Phys. Rev. D 85,
%114007 (2012).
%
%\bibitem{Janbazi}
%R. Khosravi and M. Janbazi, Phys, Rev, D 87, 016003 (2013).
%
%\bibitem{Janbazi2}
%R. Khosravi and M. Janbazi, Phys, Rev, D 89, 016001 (2014).
\end{thebibliography}
\end{document}